\documentclass[aps,pra,twocolumn,superscriptaddress]{revtex4-2}

\usepackage{graphicx}
\usepackage{dcolumn}
\usepackage{bm}
\usepackage{amsmath}
\usepackage{amssymb}
\usepackage{latexsym}
\usepackage{epsfig}
\usepackage{amsbsy}
\usepackage{array}
\usepackage{amssymb}
\usepackage{setspace}
\usepackage{bm}
\usepackage{epstopdf}
\usepackage{subfigure}
\usepackage{color}
\usepackage{siunitx}
 \usepackage{tabularray}
\usepackage{multirow}

\def\sint{\ifmmode{- \!\!\!\!\!\! \int}
    \else{\hbox{$- \!\!\!\! \int \ $}}\fi}

\begin{document}
\title{Probing minimal observable length with dark modes in an optomechanical detector}
\author{Wenlin Li}
\email{liwenlin@mail.neu.edu.cn}
\affiliation{College of Sciences, Northeastern University, Shenyang 110819, China}
\author{Xingli Li}
\affiliation{Department of Physics, The Chinese University of Hong Kong, Shatin, New Territories, Hong Kong, China}
\author{Najmeh Eshaqi-Sani}
\affiliation{Department of Mathematical, Physical and Computer Sciences, University of Parma, Parco Area delle Scienze 7/A, 43124, Parma, Italy}
\author{Wen-zhao Zhang}
\affiliation{Department of Physics, School of Physical Science and Technology, Ningbo University, Ningbo, 315211, China}
\author{Jiong Cheng}
\affiliation{Department of Physics, School of Physical Science and Technology, Ningbo University, Ningbo, 315211, China}

\date{\today}
\begin{abstract}
Several theories that attempt to unify quantum theory and gravitational theory assume that space has an observable limiting resolution related to the Planck length, denoted by $\sqrt{\beta_0}L_p$. Quantum mechanically, this concept derives a generalized uncertainty principle (GUP) and the corresponding modified commutator. The prediction and observation of GUP-induced new physics, as well as the quantitative measurement of the value of $\beta_0$, may provide substantial support for the establishment of quantum gravity theory. In this paper, we propose a comprehensive quantum framework for measuring GUP at low energy scales by utilizing the interference-induced bright-dark mode effect of oscillators in an optomechanical system. The nonlinearity induced by GUP will be amplified by the bright mode dynamics, and then be quantitatively read out by the noise spectrum of the dark mode. The measurement limit resolution of the scheme is not  constrained  by the quality factor of the oscillator. Under experimentally achievable parameters, the measurement resolution has been shown to reach $\beta_{\text{NL,lim}}=10^{-16.75}$, which is $10$ orders of magnitude lower than the electroweak level.
\end{abstract}
\maketitle
\section{Introduction}  
The unification of gravity with quantum theory has been a pivotal step in the pursuit of a grand unified theory in contemporary physics. Various approaches, including loop quantum gravity theory, string theory, and subsequent superstring theory,  incorporate the fundamental concept of the Planck length $L_p$. It is defined as the smallest resolvable scale of space, of the order of $L_p\sim $\SI[scientific-notation = fixed, fixed-exponent = -35, retain-unity-mantissa = false, exponent-product = {}]{1e-35}{\meter}~\cite{Garay1995,Hossenfelder2013}. In the event that a grand unified theory does require a restriction on the resolution of space, then, while its particular formulation remains obscure, under weak gravity case, the theory should devolve into a quantum mechanics that differs from standard quantum mechanics, since the Heisenberg uncertainty principle posits the possibility of attaining infinite precision in measuring spatial degrees of freedom by sacrificing momentum resolution  $\Delta p$. A succinct theory that guarantees a lower bound on  $\Delta q$, that is, $\Delta q_{\text{min}}=\sqrt{\beta_0}L_p$, corresponds to a modified uncertainty principle expressed as~\cite{Amati1987,Gross1988,Maggiore1993}:
\begin{equation}
\begin{split}
\Delta q\Delta p\geq \dfrac{\hbar}{2}\left[1+\beta_0\left(\dfrac{L_p\Delta p}{\hbar}\right)^2\right],
\end{split}
\label{eq:GUP}
\end{equation}
where $\beta_0$ is a dimensionless scale parameter. The relation is known as the generalized uncertainty principle (GUP), which pertains to a quantum mechanics characterized by a modified commutation relation~\cite{Garay1995,Amati1987,Pikovski2012,Bawaj2015,Bonaldi2020}:
\begin{equation}
\begin{split}
[\hat{q},\hat{p}]=i\hbar\left[1+\beta_0\left(\dfrac{L_p\hat{p}}{\hbar}\right)^2\right].
\end{split}
\label{eq:GUPc}
\end{equation}

The supernumerary correction term suggests the potential emergence of \textit{new physics} at the scale characterized by $\beta_0$~\cite{Das2008}. Specifically, when $\beta_0=1$, it appears strictly at the Planck scale. While the scale corresponding to the electroweak phenomenon is $\beta_0=10^{34}$, and the probability of $\beta_0$ exceeding this critical value is negligible, given the consistency of experimental and theoretical results at high precision on the electroweak scale over the past several decades. Consequently, the non-trivial range of $\beta_0$ is $1$ to $10^{34}$, and any phenomenon induced by $\beta_0$ in this interval can be regarded as new phenomena.

The first experiment to probe physics beyond the electroweak scale was reported by Marin by measuring the increase in the ground state energy ($E_{\text{min}}$) of a harmonic oscillator. The corresponding upper limit on $E_{\text{min}}$ was established through meticulous analysis of the first longitudinal mode of the AURIGA gravitational wave bar detector~\cite{Marin2012,Marin2014}. The experimental data constrained Planck-scale modifications to an upper limit of $\beta_0<10^{33.47}$. In addition, measurement schemes based on different physical mechanisms have also been proposed, including measuring the ground-state Lamb shift~\cite{Ali2011} and the 1S–2S level difference~\cite{Quesne2010} based on high-resolution spectroscopy of hydrogen atom; measuring the lack of violation of the equivalence principle~\cite{Ghosh2014}; investigating the symmetry breaking of the inversion operation~\cite{Pikovski2012}; measuring the nonlinear dynamics of a quantum resonator~\cite{Bawaj2015,Bonaldi2020,Li2025}; and the search for phenomena in astronomical events~\cite{Das2021}.

Among the aforementioned methods, investigating the oscillator dynamics at low energy scales is considered as an advantageous approach, as it offers the optimal resolution available, designated as $\beta_{0,\text{lim}}=10^{7.4}$.  It is  demonstrated by Ref.~\cite{Bawaj2015}  through utilizing an interferometer to observe a macroscopic oscillator (the mass is \SI[scientific-notation = fixed, fixed-exponent = -2, retain-unity-mantissa = false, exponent-product = {}]{0.01}{\gram}) operating within the classical limit . A salient benefit of this approach is its extension to the quantum regime, which is imperative given the ongoing discourse surrounding \textit{quantum} gravity effects. The cavity optomechanical system (OMS), which has been extensively studied in quantum optics, provides a complete description of an oscillator coupled to an optical field that plays the role of manipulator and detector,  via a full quantum model~\cite{Aspelmeyer2014}. Refs.~\cite{Bonaldi2020,Li2025} have discussed the possible effects of observing quantum gravity on pre-prepared, high-purity, mesoscopic mechanical oscillators via  the sideband cooling effect and has reported on theoretical speculations and experimental realizations. Generally speaking, these measurement schemes require that the oscillator be prepared into a coherent state of a considerable amplitude, and examines its non-stationary dynamics~\cite{Bawaj2015,Li2025}. The corresponding resolution is thus constrained by the oscillator's quality factor, a defect that will be expounded upon in the subsequent sections. Furthermore, the nonlinearity inherent in the OMS  will also impose constraints on the achievable resolution of the measurement~\cite{Aspelmeyer2014,Marquardt2006}. The quality factor of the oscillator and the nonlinear coupling strength can be regarded as the intrinsic properties of the system. It is challenging to enhance these two parameters without replacing the oscillator's material.

In this paper, we propose a scheme to measure the GUP parameter $\beta_0$ in an OMS comprising two oscillators that are coupled to a common optical mode~\cite{Genes2008,Sheng2020,Piergentili2021}. The measurement process is based on the stationary dynamics of the dark mode, which is formed by the interference of the two oscillators with each other, resulting in decoupling from the optical field~\cite{Genes2008,Huang2023,Lai2020,Piotrowski2023,Cao2025}. The limitations imposed by the nonlinearity of the radiation pressure interaction, resulting from such decoupling, are significantly mitigated or even eliminated.  Moreover, the stationary dynamics extend the duration over which data can be collected. In an ideal scenario,  the upper limit could extend to infinity,  thereby breaking the resolution limitation casued by quality factor. The measurement process is simulated with relevant parameters from the existing references. The data collection time exceeds that of Ref.~\cite{Li2025} by a factor of $10^4$. The measurement resolution is enhanced without the necessity of extreme levels of resonator amplitude. Subsequently, the discussion shifted to the impact of imperfect dark modes on the measurement scheme.

This paper is organized as follows: in Sec.~\ref{Noise spectrum affected by GUP} the fundamental principle of utilizing noise spectrum to analyze GUP is presented. In Sec.~\ref{GUP measurement scheme based on the coherent cancellation of the dark mode}, a thorough analysis of the designed measurement model's dynamics is conducted. Sec.~\ref{The model and system dynamics} introduces the driving effect of the input laser modulation on the bright mode, while Sec.~\ref{Excitation of bright modes and dynamical correction of dark modes} explores the amplification effect of the excited bright mode on the GUP on the dark mode noise spectrum. The actual simulation of the measurement scheme is placed in Sec.~\ref{Simulation results and data analysis}, where the measurement results in the ideal scenario are first introduced in Sec.~\ref{Ideal situation} and then analyzed in terms of the impact of parameter errors in Sec.~\ref{Actual situation}. The discussion and outlook are situated in Sec.~\ref{Discussion and conclusion}.
\section{Noise spectrum affected by GUP}  
\label{Noise spectrum affected by GUP}
\begin{figure}[]
\centering
\includegraphics[width=3.3in]{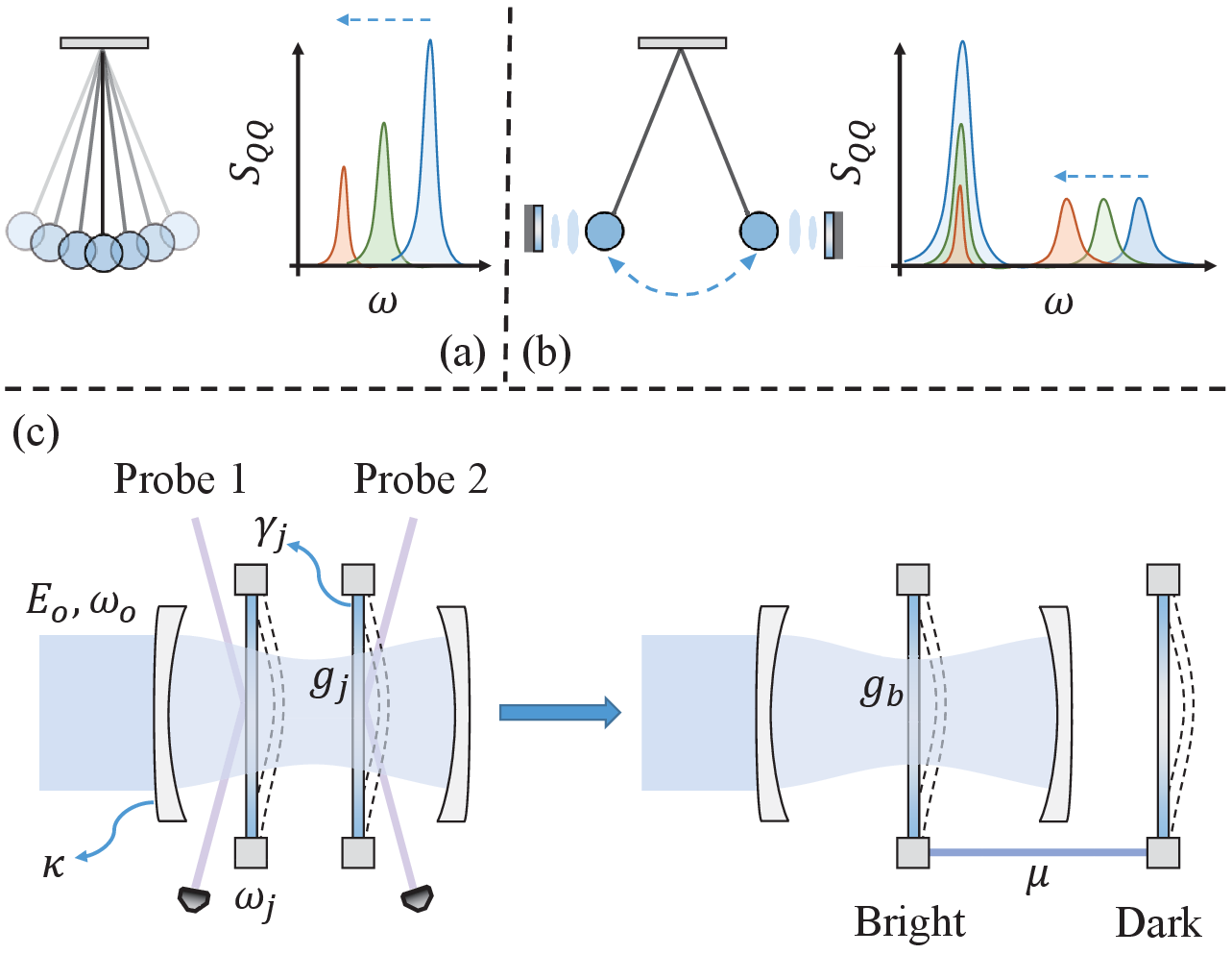}
\caption{ (a): Principle of measuring GUP by non-stationary dynamics: the oscillator undergoes free dissipation from a highly excited initial state and eventually reaches an equilibrium state. The GUP effect will cause the oscillation frequency to redshift as the amplitude dissipates. The corresponding  oscillation spectrum is measured. (b): Principle of measuring GUP by stationary dynamics: the oscillator is driven by a single-mode force, maintaining forced vibration, resulting in the corresponding fluctuation dynamics undergoing a redshift with weaker drive strength. The noise spectrum of the oscillator is measured. (c): Schematic diagram of a two-membrane OMS as the optomechanical detector in our scheme. This model can be also described by a ``bright'' mode coupled to the cavity field and a ``dark'' mode which is decoupled from the cavity field.
\label{fig:1}}
\end{figure}
We here inherit the treatment of Ref.~\cite{Bawaj2015} to deal with the modified commutation relation~\eqref{eq:GUPc}. By defining the dimensionless coordinate $\hat{q}=\sqrt{\hbar/(m\omega_b)}\hat{Q}$ and momentum $\hat{p}=\sqrt{\hbar m\omega_b}\hat{P}_G$, we can obtain the dimensionless commutation relation corresponding to GUP, which is expressed as,
\begin{equation}
\begin{split}
[\hat{Q},\hat{P}_G]=i(1+\beta_\text{NL} \hat{P}^2_G).
\end{split}
\label{eq:GUPdim}
\end{equation}
The dimensionless parameter $\beta_0$ is converted to another dimensionless parameter $\beta_{\text{NL}}=\beta_0(\hbar m\omega_b/m_p^2c^2)$ satisfying $\beta_\text{NL} \ll 1$. The modified commutation relation will degenerate into the ordinary commutation relation $[\hat{Q},\hat{P}]=i$ by introducing the transformation $\hat{P}_G=(1+\beta_{\text{NL}} \hat{P}^2/3)\hat{P}$. For a harmonic oscillator, the Hamiltonian, reformulated in terms of the operators $\hat{Q}$ and $\hat{P}$, becomes $H={\hbar\omega_b(\hat{Q}^2+\hat{P}^2)}/{2}+\hbar{\omega_b}\beta_{\text{NL}} \hat{P}^4/3$. We can then write the Hamiltonian in corresponding second quantized form as:  
\begin{equation}
\begin{split}
H=\omega_b \hat{b}^\dagger\hat{b}+\dfrac{\omega_b\beta_{\text{NL}}}{12}\left(\hat{b}^\dagger-\hat{b}\right)^4,
\end{split}
\label{eq:Hamiltonian}
\end{equation}
where $\hat{b}=(\hat{Q}+i\hat{P})/\sqrt{2}$ and $\hat{b}^\dagger=(\hat{Q}-i\hat{P})/\sqrt{2}$ are the annihilation  and creation operators conform to the standard definition and satisfy the commutation relation $[\hat{b},\hat{b}^\dagger]=1$. Assuming that the oscillator is placed in a heat reservoir with a dissipation rate $\gamma$, the corresponding quantum Langevin equation under the Heisenberg picture is~\cite{Giovannetti2001}:
\begin{equation}
\begin{split}
\dot{\hat{b}}=\left(-i\omega_b-\gamma\right)\hat{b}+i\omega_{b}\dfrac{\beta_{\text{NL}}}{3}\left(\hat{b}-\hat{b}^\dagger\right)^3+\sqrt{2\gamma}\hat{b}_{in},
\end{split}
\label{eq:mechanical oscillator QLE}
\end{equation}
where $\hat{b}_{{in}}$ is the corresponding input noise operators. It satisfies the correlation relationship of Gaussian white noise, that is, it has zero expected value and its autocorrelation function is $\langle\hat{b}^{\dagger}_{in}(t)\hat{b}_{{in}}(t')+\hat{b}_{{in}}(t)\hat{b}^{\dagger}_{in}(t')\rangle=(2\bar{n}_{b}+1)\delta(t-t')$. Here $\bar{n}_{b}=[\exp(\hbar\omega_{b}/k_bT)-1]^{-1}$ is the mean thermal excitation number for the resonator corresponding to the temperature $T$~\cite{book1}. We express the operator of the mechanical oscillator as the sum of its expectation value and its fluctuation as $\hat{b}=\langle b\rangle+\delta b$~\cite{Aspelmeyer2014,Vitali2007,Wang2014}.  The equations of motion for the classical mean field given by:
\begin{equation}
\begin{split}
\langle\dot{{b}}\rangle&=\left(-i\omega_b-\gamma\right)\langle{b}\rangle+i\omega_{b}\dfrac{\beta_{\text{NL}}}{3}\left(\langle{b}\rangle-\langle{b}\rangle^*\right)^3,
\\&\simeq\left(-i\omega_{b,\langle b\rangle}-\gamma\right)\langle{b}\rangle,
\end{split}
\label{eq:mechanical oscillator QLE}
\end{equation}
where $\omega_{b,\langle b\rangle}=\omega_b(1+{\beta_{\text{NL}}}\vert \langle b\rangle\vert^2)$ is a GUP-induced modified oscillator frequency that depends on the expectation value $\vert\langle b\rangle\vert$. 

When the oscillator is prepared with a large coherent amplitude, the GUP becomes observable in the frequency domain. This concept has been employed in several reported experiments~\cite{Bawaj2015,Bonaldi2020}. However, it must be noted that the dissipation rate results in the amplitude decaying by a factor of $\exp(-\gamma t)$. Consequently, the frequency correction is only applicable within a constrained time interval  $t\sim k\gamma^{-1}$. In order to observe the frequency correction induced by the GUP, the most optimistic analysis allows us to collect data of the oscillator dynamics up to $k\sim 2$; beyond this point, the signal will dissipate and become too small to observe. In instances where a quantitative evaluation is necessary, the corresponding interval is reduced to $k<10^{-2}$~\cite{Li2025}. It is attributed to the necessity of maintaining constant amplitude over the specified time interval to ensure that $\omega_{b,\langle b\rangle}$ remains constant. The length of the time domain determines the resolution of the frequency domain, so the measurement resolution of parameters $\beta_{\text{NL}}$ and $\beta_0$, denote as  $\beta_{\text{NL,lim}}$ and  $\beta_{0,\text{lim}}$ satisfy the following relationships:
\begin{equation}
\begin{split}
\beta_{\text{NL,lim}}=\dfrac{k}{\vert A_{\text{lim}}\vert^2\text{Q}},\,\,\,\,\,\, \beta_{0,\text{lim}}=\dfrac{k m_p^2c^2}{\vert A_{\text{lim}}\vert^2\text{Q}\hbar m\omega_b},
\end{split}
\label{eq:resolution}
\end{equation}
where $\vert A_{\text{lim}}\vert$ represents the maximum amplitude of the oscillator that can be prepared, and $\text{Q}=\omega_b/\gamma$ is the quality factor of the oscillator. In the event that these two quantities are constant, a natural idea to overcoming this detection limit would be to provide the resonator with a continuous drive, thereby preventing it from decaying.  In an ideal scenario, the data acquisition time is infinite, corresponding to an infinitesimal frequency domain resolution. However, the drive will result in the dynamics of the oscillator being actually a forced oscillation. Its corresponding oscillation frequency is therefore the drive frequency rather than the eigenfrequency itself~\cite{Mari2009}. Consequently, the GUP signal cannot be read.

In contrast to the coherent oscillation spectrum, which has been the focus of previous studies, our investigation focuses on the resonator noise spectrum. Under the condition of $\beta_\text{NL}\ll 1$, the fluctuation operator equation is linearized by the mean field approximation as follows:
\begin{equation}
\begin{split}
\dot{\delta{b}}=&\left(-i\omega_b-\gamma\right)\delta{b}\\&-i4\omega_{b}{\beta_{\text{NL}}}\left[\text{Im}(\langle b\rangle)\right]^2\left(\delta{b}-\delta{b}^\dagger\right)+\sqrt{2\gamma}\hat{b}_{in},\\
\simeq& \left(-i\omega'_{b,\langle b\rangle}-\gamma\right)\delta{b}+\sqrt{2\gamma}\hat{b}_{in},
\end{split}
\label{eq:mechanical oscillator QLE}
\end{equation}
where $\omega'_{b,\langle b\rangle}=\omega_b(1+4{\beta_{\text{NL}}}\left[\text{Im}(\langle b\rangle)\right]^2)\simeq \omega_{b,\langle b\rangle}$. The simplified expression above neglects the higher-order terms of the fluctuation operator and the terms related to the conjugate operator $\delta b^\dagger$ due to the linearization approximation and the rotational-wave approximation (RWA). 

Introducing the Fourier domain operators
\begin{equation}
\begin{split}
{\delta b}[\omega]=\dfrac{1}{\sqrt{2\pi}}\int dt{\delta b}[t]e^{-i\omega t},
\end{split}
\label{eq:Fourier}
\end{equation}
the Langevin equation becomes:
\begin{equation}
\begin{split}
-i\omega\delta b[\omega]=\left(-i\omega_{b,\langle b\rangle}-\gamma\right)\delta{b}[\omega]+\sqrt{2\gamma}\hat{b}_{in}[\omega],
\end{split}
\label{eq:mechanical oscillator QLE fey}
\end{equation}
and the corresponding symmetrized noise spectrum can be expressed as:
\begin{equation}
\begin{split}
\bar{S}_{QQ}[\omega]=&\dfrac{1}{2}(S_{QQ}[\omega]+S_{QQ}[-\omega]),\\
=&\dfrac{\gamma (n_b+\frac{1}{2})}{(\vert\omega\vert-\omega_{b,\langle b \rangle})^2+\gamma^2},
\end{split}
\label{eq:Noise spectrum}
\end{equation}
where $S_{QQ}[\omega]=\int dt e^{i\omega t}\langle {\delta Q}(t){\delta Q}(0)\rangle$ is the quantum noise spectral density with the coordinate fluctuation operator is defined as $\delta Q=(\delta b^\dagger+\delta b)/\sqrt{2}$. Equation~\eqref{eq:Noise spectrum} indicates that  the noise spectrum can also be regarded as a pointer, with the peak of the spectrum located at the position of $\omega_{b,\langle b\rangle}$. In contradistinction to the classical spectrum, the coherent drive does not force the noise spectrum to resonate with it. Consequently, the noise spectrum can be measured while the coherent drive continues to maintain the amplitude of the oscillator without dissipation. As previously indicated, the frequency resolution limitation of the data acquisition time interval, resulting from oscillator dissipation, can be circumvented. However, new limitations have emerged. A substantial GUP-induced observable translation necessitates a coherent state of the oscillator with a considerable amplitude, thereby corresponding to a remarkably elevated classical peak in the spectrum. The position of this peak aligns with the frequency of the coherent drive. Consequently, in instances where both the drive and the oscillator are resonated, the classical peak will submerge the the fluctuation peak, rendering  it invisible. In reality, given the substantial disparity in size between the classical peak and the fluctuation peak, the distance between these two in the frequency domain must be adjusted  far enough, which is a significant detuning between the drive field and the oscillator. This will result in a significant reduction in the factor $\vert A_{\text{lim}}\vert$ in Eq.~\eqref{eq:resolution} when the drive intensity is held. Furthermore, continuous driving can induce additional dynamical effects in the oscillator, such as frequency renormalization, potentially introducing  additional noise or spurious signals that render the GUP undetectable.
\section{GUP measurement scheme based on the coherent cancellation of the dark mode}  
\label{GUP measurement scheme based on the coherent cancellation of the dark mode}
In this section, the previously mentioned measurement restriction caused by drive frequency detuning is optimized within the designed optomechanical system. It is noteworthy that, under specific conditions, the supermodes of a three-body system can be considered as either bright or dark modes, which corresponds to coherent constructive and destructive effects, respectively. We subsequently demonstrate that the noise spectrum of the dark mode, influenced by the GUP, will also generate a frequency shift. In contrast to the behavior of a single oscillator, the shift distance of the spectrum peak is not only contingent on the coherent oscillation amplitude of the dark mode, but also determined by that of the bright mode. Consequently, the drive can be continuously applied to the bright mode, and its coherent oscillation spectrum will not mask the noise spectrum of the dark mode.
\subsection{The model and system dynamics}
\label{The model and system dynamics}
As shown in Fig.~1, the measurement scheme under consideration is based on two mechanical resonators  that are coupled to a common cavity. This model, as a typical structure of a multi-mode OMS, has been implemented and reported by multiple experimental groups~\cite{Piergentili2021,Sheng2020,Piergentili2018,Wu2023,Lake2020}. The laser beam reflection from each  resonator is used to directly monitor the resonator's motion, with a sensitivity that can reach $10^{-14}$\,m/$\sqrt{\text{Hz}}$~\cite{Sheng2020}. This level of accuracy is sufficient to obtain the corresponding noise spectrum.  After ignoring the detection lasers, the system Hamiltonian is expressed as~\cite{Li2020,Piergentili2021}:
\begin{equation}
\begin{split}
H_s/\hbar=&\omega_c \hat{a}^\dagger\hat{a}+\sum_{j=1,2}\left[\omega_{bj}\hat{b}_j^\dagger\hat{b}_j-g_j\hat{a}^\dagger\hat{a}\left(\hat{b}_j^\dagger+\hat{b}_j\right)\right
]\\&+\sum_{o=h,c}iE_o(\hat{a}^\dagger e^{-i\omega_{o} t}-\hat{a}e^{i\omega_{o} t}),
\end{split}
\label{eq:system Hamilton}
\end{equation}
and 
\begin{equation}
\begin{split}
H_{g}/\hbar=\sum_{j=1,2}\left[\dfrac{\omega_{bj}\beta_{\text{NL},j}}{12}\left(\hat{b}^\dagger_j-\hat{b}_j\right)^4\right].
\end{split}
\label{eq:system Hamilton}
\end{equation}
Here $\hat{a}$ and $\hat{b}_j$ are the optical and mechanical annihilation operators, $\omega_c$ and $\omega_{bj}$ are the resonance frequencies of the cavity field and the $j$-th mechanical resonator, while $g_j$ is the corresponding single-photon optomechanical coupling rate. The drive is comprised of a modulation consisting of frequencies $\omega_o$ with a corresponding modulation intensity $E_o$ satisfying $E_o=\sqrt{2\kappa_\text{in}P_o/\hbar\omega_o}$, where $\kappa_\text{in}$ is the cavity field decay rate through the input port and $P_o$ is the laser input power~\cite{Mari2009}. The cavity mode and the mechanical resonators are coupled to their corresponding thermal reservoir through fluctuation-dissipation processes, which are described in the Heisenberg picture by adding dissipative and noise terms, yielding the following quantum Langevin equations, after adopting a rotating framework with respect to the modulation frequency $\omega_c$:
\begin{equation}
\begin{split}
\dot{\hat{a}}=&\left\{i\left[-\Delta_1+\sum_jg_j\left(\hat{b}_j+\hat{b}^\dagger_j\right)\right]-\kappa\right\} \hat{a}\\&+E_he^{-i\Delta_2t}+E_c+\sqrt{2\kappa}\hat{a}_{in},\\
\dot{\hat{b}}_j=&\left(-i\omega_{bj}-\gamma_j\right)\hat{b}_j+i\omega_{bj}\dfrac{\beta_{\text{NL}}}{3} \left(\hat{b}_j-\hat{b}_j^\dagger\right)^3\\&+ig_j \hat{a}^\dagger \hat{a} +\sqrt{2\gamma}\hat{b}_{in,j}.
\end{split}
\label{eq:QLE}
\end{equation}
Here $\kappa$ is the total optical loss rate greater than $\kappa_\text{in}$, and $\gamma_j$ is the mechanical amplitude decay rate of the $j$-th oscillator. $\hat{a}_{{in}}$ and $\hat{b}_{{in,j}}$ are the corresponding input noise operators. They satisfy the correlation relationship of Gaussian white noise, that is, they have zero expected value and are all uncorrelated with each other. Their autocorrelation function are $\langle\hat{a}^{\dagger}_{in}(t)\hat{a}_{{in}}(t')+\hat{a}_{{in}}(t)\hat{a}^\dagger_{{in}}(t')\rangle=\delta(t-t')$ and $\langle\hat{b}^{\dagger}_{in}(t)\hat{b}_{{in}}(t')+\hat{b}_{{in}}(t)\hat{b}^\dagger_{{in}}(t')\rangle=(2\bar{n}_{b,j}+1)\delta_{jj'}\delta(t-t')$. Here $\bar{n}_{b,j}=[\exp(\hbar\omega_{bj}/k_bT)-1]^{-1}$  is the mean thermal excitation number for the $j$-th resonator corresponding to the reservoir temperature $T$~\cite{book1}.

In the subsequent section of this paper, it is constrained that all the parameters corresponding to the measurement schemes to be discussed satisfy the following two principles. The first is that the radiation pressure intensity satisfies the weak coupling regime, that is, both $g_j /\omega_{bj}$ and $g_j /\kappa$ are always satisfied. The second principle requires a scenario wherein the mechanical resonator exhibits a non-zero coherent amplitude, attributable to the excitation imparted by the modulated pump, yet the all eigenvalues of the corresponding Jacobian matrix possess negative real parts for the entirety of the periodic cycle, thereby precluding self-sustaining dynamics, can be attained through meticulous modulation of the input strength of the pump laser $E_o$ to ensure the maintenance of a stable condition~\cite{Mari2009,Farace2012}. In this case, the optomechanics is essentially in its semiclassical limit and can be approximated  as a Gaussian system~\cite{Weedbrook2012,Verlot2009}. The quantum Langevin equations become a set of coupled classical deterministic equations for the corresponding optical and mechanical complex variables  $a$ and $b$, as follows~~\cite{book2,Li2025,Weiss2016}: 
\begin{equation}
\begin{split}
\dot{{a}}=&\left\{i\left[-\Delta_1+\sum_jg_j\left({b}_j+b^*_j\right)\right]-\kappa\right\} {a}\\&+E_he^{-i\Delta_2t}+E_c+\sqrt{2\kappa}a_{in},\\
\dot{{b}}_j=&\left(-i\omega_{bj}-\gamma_j\right){b}_j+i\omega_{bj}\dfrac{\beta_{\text{NL}}}{3} \left({b}_j-{b}_j^*\right)^3\\&+ig_j \vert{a}\vert^2 +\sqrt{2\gamma}{b}_{in,j}.
\end{split}
\label{eq:CLE}
\end{equation}
Here $a_{in}$ and $b_{in}$ are already classic $c$-number noises, which satisfy the modified autocorrelation functions $\langle{a}^{*}_{in}(t){a}_{{in}}(t')\rangle=\delta(t-t')/2$ and $\langle{b}^{*}_{in}(t){b}_{{in}}(t')\rangle=(\bar{n}_{b,j}+1/2)\delta_{jj'}\delta(t-t')$. This is due to the fact that the $c$-number no longer exhibits the commutative relationship~\cite{Li2020}.

\subsection{Excitation of bright modes and dynamical correction of dark modes}
\label{Excitation of bright modes and dynamical correction of dark modes}
We then conveniently rewrite the forced vibration dynamics of the resonator as~\cite{Li2020,Mari2009}
\begin{equation}
b_j(t)=\beta_{j,0}+A_je^{-i\Delta_2 t},
\label{eq:ansatz b}
\end{equation}
where $\beta_j $ are constant terms describing the new equilibrium position of the oscillator pushed by the radiation pressure, $A_j$ are slowly varying complex amplitudes of the resonator. The driving frequency is selected to be the standard frequency of the reference frame, which satisfying $\Delta_2\simeq \bar{\omega}_b:=(\omega_{b1}+\omega_{b2})/2$. Inserting Eq.~\eqref{eq:ansatz b} into Eq.~\eqref{eq:CLE}, and solving it formally by neglecting the transient term related to the initial values, we have:
\begin{equation}
\begin{split}
a(t)=&\int_0^tdt'\left\{e^{\mathcal{L}(t-t')}[E_he^{-i\Delta_2t'}+E_c]\right.\\&\times\left.\exp\left[2i g_q \int_{t'}^t dt'' \vert A_b(t'')\vert\cos(\bar{\omega}t''-\theta)\right] \right\},
\end{split}
\label{eq:cavity field pump}
\end{equation}
where $\mathcal{L}=-i\Delta_1-\kappa$. $A_b$ is a supermodel that we define here, expressed as:
\begin{equation}
\begin{split}
A_b=\dfrac{g_1A_1+g_2A_2}{\sqrt{g_1^2+g_2^2}}.
\end{split}
\label{eq:supermodel}
\end{equation}
As revealed by Eq.~\eqref{eq:cavity field pump}, $A_b$ is the sole mode that directly couples with the cavity field with an effective coupling coefficient $g_b=\sqrt{g_1^2+g_2^2}$. Consequently, it corresponds to a ``bright" mode. When both resonators have higher quality factors, the amplitude $A_b(t)$ is much slower than the fast oscillations at $\Delta_2$. Thus, it can be regarded as a constant in the integration with the index $t''$~\cite{Marquardt2006}. Then Eq.~\eqref{eq:cavity field pump} can be simplified as:
\begin{equation}
\begin{split}
a(t)=&e^{i\psi_b(t)}\\&\times\int_0^tdt' e^{\mathcal{L}(t-t')}\left[E_he^{-i\Delta_2t'}+E_c\right]e^{-i\psi_b(t')},
\end{split}
\label{eq:cavity field pump2}
\end{equation}
where $\psi_b (t)=\xi \sin(\Delta_2 t-\theta)$ with $\xi=2g_b\vert A_b\vert /\Delta_2$.
The exponential term $e^{-i\psi_b(t')}$ in the integral can be extended by the Jacobi-Anger expansion, that is, $(e^{-i\xi\sin \phi}=\sum_n J_n(-\xi)e^{in\phi}$, $J_n$  is the $n$-th Bessel function of the first kind), and after neglecting a quickly decaying term, we finally get
\begin{equation}
\begin{split}
a(t)=&e^{i\psi_b(t)}\left[E_h\sum_{n=-\infty}^{\infty}\dfrac{J_n\left(-\xi\right)e^{in(\Delta_2t-\theta)-i\Delta_2 t}}{in\Delta_2-\mathcal{L}_h}\right.\\&\left.+E_c \sum_{n=-\infty}^{\infty}\dfrac{J_n\left(-\xi\right)e^{in(\Delta_2t-\theta)}}{in\Delta_2-\mathcal{L}_c}\right],
\end{split}
\label{eq:cavity field solution}
\end{equation}
where $\mathcal{L}_h=-i(\omega_c-\omega_h)-\kappa$, and  $\mathcal{L}_c=-i\omega_c-\kappa$.  We re-express it as a sum of a series of sidebands: $\vert a\vert^2=\sum_n P_ne^{in\Delta_2 t}$. Here we neglect all terms oscillating faster than $\Delta_2$ meaning that only the resonant terms with the first-order sideband $e^{-i\Delta_2 t}$ is kept, with the coefficient 
\begin{equation}
\begin{split}
P_{-1}=&E_h^2\sum_{n=-\infty}^{\infty}\dfrac{J_n\left(-\xi\right)J_{n+1}\left(-\xi\right)e^{i\theta }}{[in\Delta_2-\mathcal{L}_h][-i(n+1)\Delta_2-\mathcal{L}^*_h]}\\
&+E_c^2\sum_{n=-\infty}^{\infty}\dfrac{J_n\left(-\xi\right)J_{n+1}\left(-\xi\right)e^{i\theta }}{[in\Delta_2-\mathcal{L}_c][-i(n+1)\Delta_2-\mathcal{L}^*_c]}\\
&+E_hE_c\sum_{n=-\infty}^{\infty}\dfrac{J_n\left(-\xi\right)^2}{[in\Delta_2-\mathcal{L}_h][-in\Delta_2-\mathcal{L}^*_c]}\\
&+E_hE_c\sum_{n=-\infty}^{\infty}\dfrac{J_n\left(-\xi\right)J_{n+2}\left(-\xi\right)e^{2i\theta}}{[in\Delta_2-\mathcal{L}^*_h][-i(n+2)\Delta_2-\mathcal{L}_c]},
\end{split}
\label{eq:cavity field solution}
\end{equation}
 By defining the following dimensionless auxiliary functions $\mathcal{F}_{1,2,3,4}(\vert A_b\vert, \Delta_2, \kappa,\omega_h,\omega_c,g_q)$ as:
 \begin{equation}
\begin{split}
\mathcal{F}_1=&\dfrac{E_h^2}{\vert A_b\vert}\sum_{n=-\infty}^{\infty}\dfrac{J_n\left(-\xi\right)J_{n+1}\left(-\xi\right)}{[in\Delta_2-\mathcal{L}_h][-i(n+1)\Delta_2-\mathcal{L}^*_h]}\\&+\dfrac{E_c^2}{\vert A_b\vert}\sum_{n=-\infty}^{\infty}\dfrac{J_n\left(-\xi\right)J_{n+1}\left(-\xi\right)}{[in\Delta_2-\mathcal{L}_c][-i(n+1)\Delta_2-\mathcal{L}^*_c]},
\end{split}
\label{eq:cavity field solution}
\end{equation}
and 
\begin{equation}
\begin{split}
\mathcal{F}_2=&E_1E_2\sum_{n=-\infty}^{\infty}\dfrac{J_n\left(-\xi\right)J_{n}\left(-\xi\right)}{[in\Delta_2-\mathcal{L}_h][-in\Delta_2-\mathcal{L}^*_c]},
\end{split}
\label{eq:cavity field solution}
\end{equation}
and 
\begin{equation}
\begin{split}
\mathcal{F}_3=\dfrac{E_1E_2}{\vert A_b\vert^2}\sum_{n=-\infty}^{\infty}\dfrac{J_n\left(-\xi\right)J_{n+2}\left(-\xi\right)e^{2i\theta}}{[in\Delta_2-\mathcal{L}^*_h][-i(n+2)\Delta_2-\mathcal{L}_c]},
\end{split}
\label{eq:cavity field solution}
\end{equation}
In the frame rotating at the fast reference frequency $\Delta_2$, we can write the slowly varying amplitudes equation in a more compact form:
\begin{equation}
\begin{split}
\dot{A}_j(t)=&\left[-\gamma_j-i\Delta\omega_j\right] A_j(t)+\sqrt{2\gamma_j}b_{in,j}\\&+ig_j[A_b(t) \mathcal{F}_1+\mathcal{F}_2+A_b(t)^2\mathcal{F}_3]\\&+i\omega_{bj}\beta_{\text{NL},j}\vert A_j\vert^2 A_j,
\label{eq:first-order finnal}
\end{split}
\end{equation}
where $\Delta \omega_j=\omega_j-\Delta_2$ is the driving-eigenfrequency detuning.

We consider a simplified case corresponding to $g_1=g_2=g$, which can always be achieved in an actual measurement process by adjusting the position of two membranes (such as fixing one  membrane and continuously slowly moving the other one within a range~\citep{Piergentili2018}).  We also assume $\beta_{\text{NL,1}}=\beta_{\text{NL,2}}=\beta_{\text{NL}}$ because the difference between the two resonators is actually very subtle. These two conditions can help us avoid some tedious algebraic calculations, while any slight damage to them does not have a practical impact on the detection scheme. We will explain this issue in the subsequent numerical simulations. After defining the supermode conjugated with $A_b$ as
\begin{equation}
\begin{split}
A_d=\dfrac{g_1A_2-g_2A_1}{\sqrt{g_1^2+g^2_2}}=\dfrac{A_2-A_1}{\sqrt{2}},
\label{eq:d mode}
\end{split}
\end{equation}
and substituting the corresponding input noise $b_{in,b(d)}$, the equations for amplitude variables of the two supermodes under the above conditions respectively are: 
\begin{equation}
\begin{split}
\dot{A}_b(t)=&-\Gamma_b A_b(t)+b_{in,b}\\&+i\sqrt{2}g[A_b(t) \mathcal{F}_1+\mathcal{F}_2+A_b(t)^2\mathcal{F}_3]-\mu A_d\\&+\dfrac{\bar{\omega}_{b}\beta_\text{NL}}{2}(2\vert A_b\vert^2A_d+A_b^2A_d^*+\vert A_d\vert^2A_d),
\label{eq:bright mode}
\end{split}
\end{equation}
and 
\begin{equation}
\begin{split}
\dot{A}_d(t)=&-\Gamma_d A_d-\mu A_b+b_{in,d}\\&+\dfrac{\bar{\omega}_{b}\beta_\text{NL}}{2}(2\vert A_b\vert^2A_d+A_b^2A_d^*+\vert A_d\vert^2A_d),
\label{eq:dark mode}
\end{split}
\end{equation}
with the effective eigenfrequency and dissipation rate
\begin{equation}
\begin{split}
\Gamma_b=\Gamma_d=i\left[\dfrac{\Delta \omega_{m1}+\Delta \omega_{m2}}{2}\right]+\dfrac{\gamma_1+\gamma_2}{2},
\label{eq:Gammab}
\end{split}
\end{equation}
and an effective beam splitter coupling with the strength:
\begin{equation}
\begin{split}
\mu=i\left[\dfrac{ \omega_{m2}- \omega_{m1}}{2}\right]+\dfrac{\gamma_2-\gamma_1}{2}.
\label{eq:mu}
\end{split}
\end{equation}

Equation~\eqref{eq:bright mode} elucidates the impact of the cavity field on the bright mode. $\mathcal{F}_1$ and $\mathcal{F}_3$ delineate the tiny frequency modification and compression induced by the cavity field. $\mathcal{F}_2$ characterizes the coherent driving provided by the cavity field, and its strength is contingent on $E_1E_2$, corresponding to the interference between the two drivers. This effect is stronger than the preceding ones, owing to the use of Bessel functions with matching indices, enabling the bright mode to attain a coherent state with substantial amplitude. Additionally, the nonlinear Bessel function predicts a saturation effect, wherein large oscillator amplitudes resist further enhancement despite increased pump laser intensity.

It is noteworthy that Eq.~\eqref{eq:bright mode} can also describe the dynamics of a single resonator if all terms relating to $A_d$ are neglected. Therefore, as previously discussed, if the drive is  maintained, the oscillation frequency of  either a single resonator or the bright mode is actually $\Delta_2$, which cannot be used to measure GUP.  With regard to the noise spectrum, under the assumption that it will not be submerged, meticulous calibration of the intensity and frequency of two pump lasers is still necessary until $\mathcal{F}_1$ is a real function. Otherwise,  cavity-induced and  GUP-induced frequency correction is indistinguishable from each other, while the former is a spurious signal.

Equation~\eqref{eq:dark mode} indicates that mode $A_d$ is decoupled from the optical field and can only be indirectly excited by mode $A_b$. We therefore define it as a ``dark" mode. In an ideal scenario where $\omega_1=\omega_2$, this indirect excitation disappears, resulting in $\langle A_d\rangle\sim 0$. Nevertheless, the GUP-induced frequency correction in the dark mode depends not only on its own coherent amplitude $A_d$, but also on the amplitude of the bright mode $A_b$. Since the dark mode is never driven by external forces, we can still quantitatively measure the GUP through its noise spectrum, even if the bright mode is continuously driven. In addition, the dark mode is unaffected by the radiation pressure interaction; thus, precise modulation of the driving field is unnecessary.

\section{Simulation results and data analysis}
\label{Simulation results and data analysis}
The specific measurement steps of the proposed scheme are as follows: Two pump laser frequencies are injected into the cavity. Once the system has stabilized, the long-term dynamics of the two resonators are respectively measured and collected. The corresponding bright mode amplitude $\langle A_b\rangle$ and the dark mode noise spectrum are calculated, and the position of the noise spectrum peak, $\omega'_b$, is subsequently recorded. Subsequent to the modification of the pump laser power and the reiteration of the aforementioned steps, a series of scattered points $(\vert\langle A_b\rangle\vert^2, \omega'_b)$ is obtained. A linear fitting is then performed on the aforementioned scattered points, and the estimated value of $\beta_{\text{NL}}$, denoted as $\beta'_{\text{NL}}$, can be calculated based on the slope $k$ with the corresponding quantitative relationship: 
\begin{equation}
\begin{split}
\beta'_{\text{NL}}=\dfrac{k}{\bar{\omega}_b}.
\label{eq:slope}
\end{split}
\end{equation}
Furthermore, the numerical study is based on a realistic scenario, which is the most pertinent for applications, and the set of parameters from Ref.~\cite{Bonaldi2020} is taken into consideration, that is, $\bar{\omega}_{b}/2\pi=$\SI{525}{kHz}, $\delta=\vert\omega_{b1}-\omega_{b2}\vert/2\leq $\SI{5}{Hz}, $\text{Q}_j=\omega_{bj}/\gamma_j\in[10^7,10^6]$, $g_j/2\pi=$\SI{1}{Hz}, $\kappa=$\SI{2.2}{MHz}, $\kappa_{\text{in}}=\kappa/2$, and $P_{c,h}\leq$\SI{100}{\micro\watt}. The wavelength of the laser is selected as $1064$\,nm.

We will then first analyze the ideal case, i.e., two resonators are identical and correspond to perfect dark modes. Then, we will proceed to analyze how discrepancies between resonators affect measurement schemes.
\subsection{Ideal situation: two identical resonators}
\label{Ideal situation}
\begin{figure}[]
\centering
\includegraphics[width=3in]{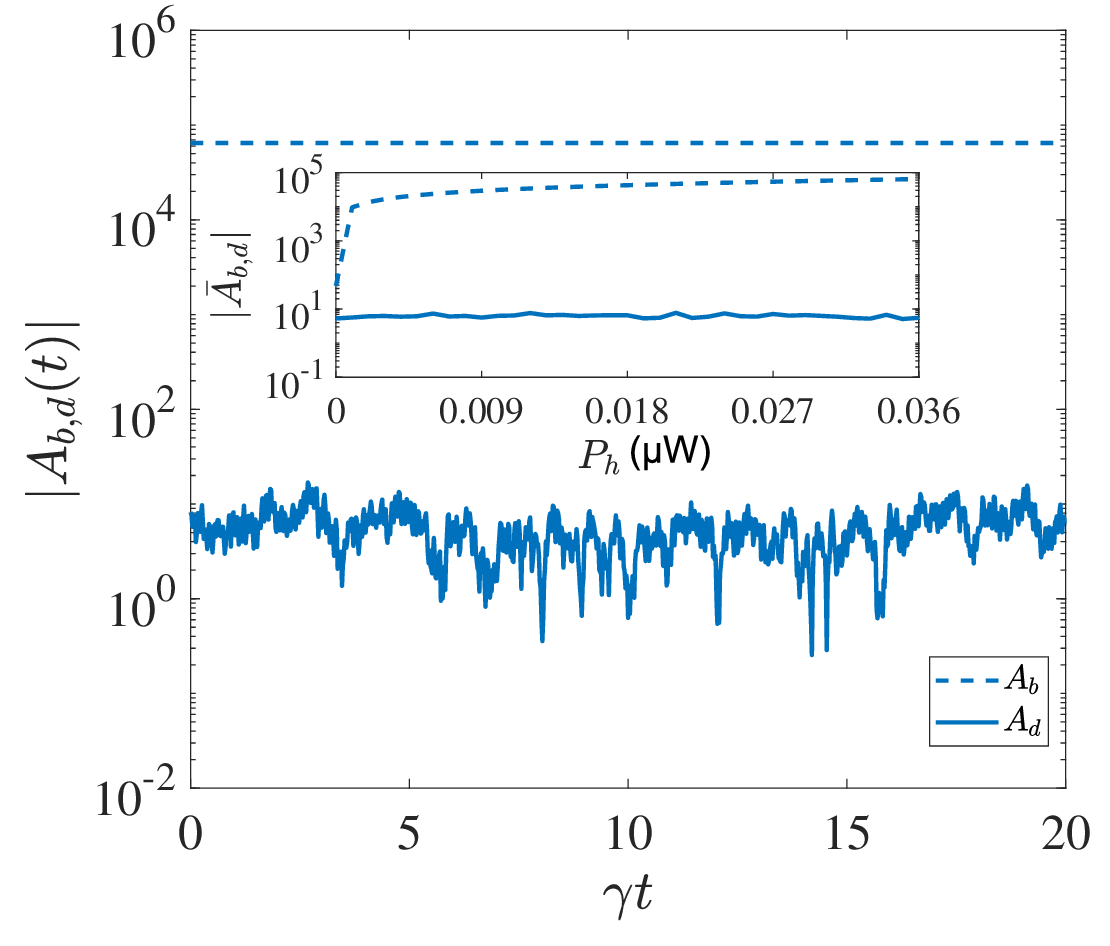}
\caption{ The time-evolved amplitude of the bright mode and dark mode. The inset illustrates the average amplitude of the bright and dark mode, i.e., $\vert\bar{A}_{b,d}\vert=\lim_{\tau\rightarrow\infty}\int_0^\tau\vert{A}_{b,d}(t)\vert dt/\tau$, as a function of one driving laser intensity $P_h$ in the case that $P_c=$\SI{100}{\micro\watt} is fixed. The dimensionless parameters corresponding to those given in the main text are: $\Delta_1=1.2857$, $\Delta_2=1$, $\kappa=4.1905$, $\gamma_1=\gamma_2=10^{-7}$, $g=1.9048\times 10^{-6}$, $\delta=0$ and $\beta_{\text{NL}}=0$ under the unit $\bar{\omega}_b=1$. $T=$\SI{0.1}{mK} so that $\bar{n}_{b,1}=\bar{n}_{b,2}\simeq 40$. The main figure corresponds to the case $P_h=$\SI{0.036}{\micro\watt}. 
\label{fig:2}}
\end{figure}

\begin{figure}[]
\centering
\includegraphics[width=3in]{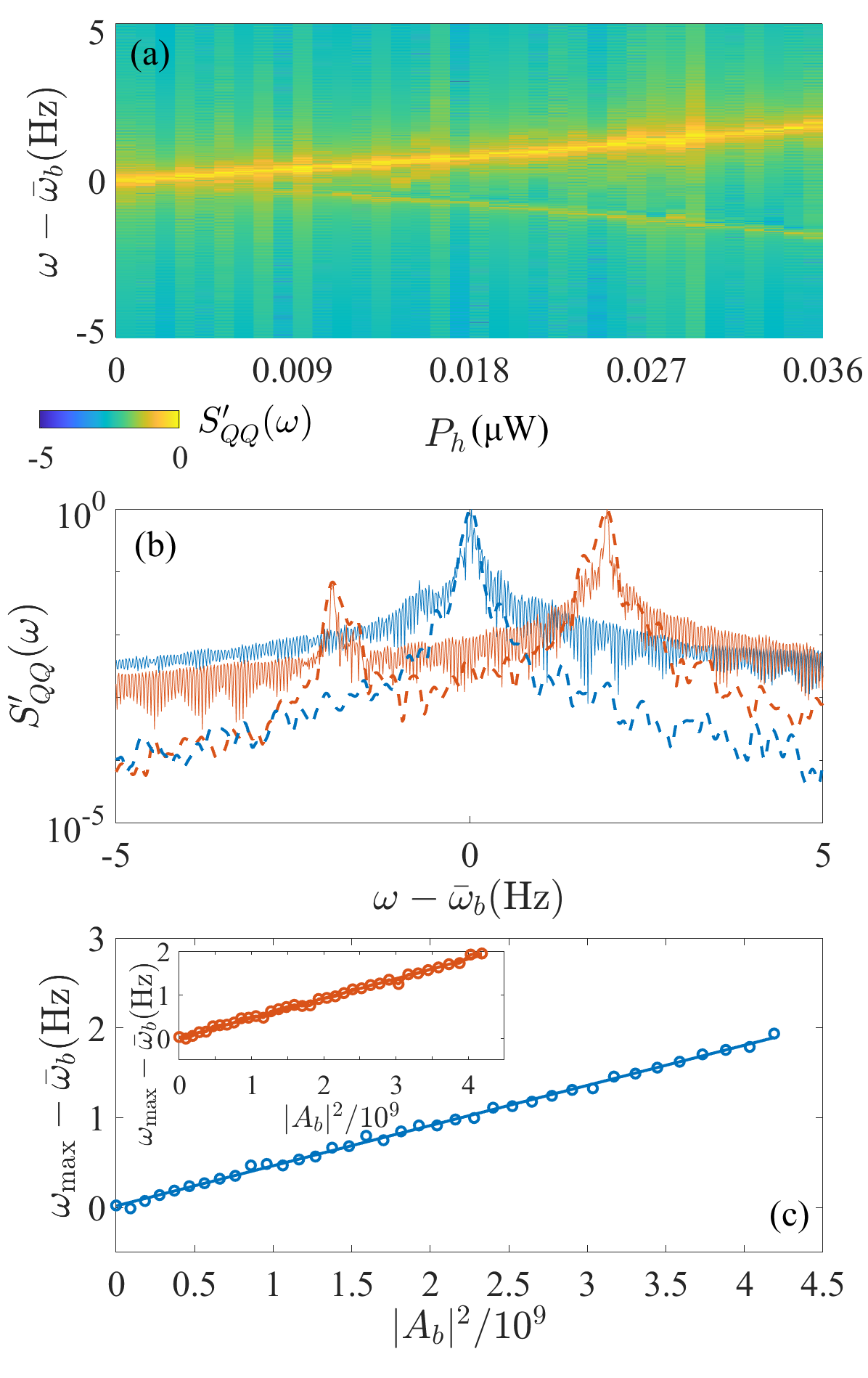}
\caption{(a): The normalized noise spectrum of the dark mode in the frequency domain with varied driving intensity $P_h$. The spectrum is obtained by directly performing the Fourier transform on the autocorrelation function. Normalization here means that the spectrum corresponding to each value of $P_h$ is divided by its respective peak value, as it is discussed in the main text. (b): The blue (red) solid line is the normalized noise spectrum corresponding to $P_h=$\SI{0}{\micro\watt} (\SI{0.036}{\micro\watt}). The  corresponding dotted lines illustrate the noise spectra obtained after processing using the Welch's method. (c): The scattered points represent the position of the noise spectrum peak obtained by the Welch's method as a function of $\vert A_b\vert ^2$, and the solid line is the corresponding linear fitting result. The inset shows the corresponding result obtained by the Fourier transformation on the autocorrelation function. Here we take $\beta_{\text{NL}}=10^{-15}$. The Welch's method employs a Blackman window with an length of $ \gamma t = 5$ and an overlap of $\gamma t = 2$. The other parameters are the same as those used in Fig.~\ref{fig:2}.
\label{fig:3}}
\end{figure}
Under the condition that the two resonators are identical, that is, $\omega_{b1}=\omega_{b2}$, $\gamma_1=\gamma_2$, we have $\Gamma_b=\Gamma_d=\gamma$ and $\mu=0$. At this time, the dark mode will not be excited at all, while due to the resonance condition, the bright mode can be driven to a larger amplitude with smaller laser powers. We simulate Eq.~\eqref{eq:CLE} up to $\gamma t=23$. The initial state of the cavity field is designated as the vacuum state, while the resonator initial states  are thermal states. We assume that the system has achieved a steady state following $\gamma t=3$. Indeed, for transient processes, this duration is already substantial. The subsequent interval of $\gamma t=20$ is recorded and utilized for data analysis. In Fig.~\ref{fig:2}, the amplitude evolution of the bright and dark modes is illustrated over the course of this time interval. In the inset of Fig.~\ref{fig:2}, we plot the mean value of the amplitude over time in the interval as a function of $P_h$ with fixed $P_c$. It indicates that a drive of the order of  $P_{c,h}\leq$\SI{0.01}{\micro\watt} can excite the dimensionless oscillator amplitude to a level that is approximately $10^5$. Correspondingly, the dark mode is never excited. The saturation effect is also apparent here. It shows that $\vert\langle\bar{A}_b\rangle\vert$ increases rapidly with drive strength in the low-power regime but plateaus for $P_h>$\SI{0.005}{\micro\watt}. This saturation implies that blindly increases in driving power provide limited improvement in resolution and may induce system instability.

Figure~\ref{fig:3} illustrates the noise spectrum of the dark mode within the frequency-power plane. In order to enhance clarity, we select $\beta_\text{NL}=10^{-15}$, corresponding to a pronounced GUP-induced frequency shift. The spectra are directly obtained via Fourier transformation of the autocorrelation function. Here each discrete power value $P_h$ corresponds to a spectrum that is normalized separately relative to its own peak. As illustrated in Fig.~\ref{fig:3}(b), the spectrum corresponds to the maximum and minimum values of the power range selected in Fig.~\ref{fig:3}(a). Quantum and thermal noise have been observed to cause significant fluctuations, which have obscured the precise peak positions. To mitigate this, we implement the Welch's method, sacrificing spectral resolution for improved smoothness. The original and Welch-processed spectra are depicted by the solid and dashed lines in Fig.~\ref{fig:3}(b), respectively. The peak frequencies extracted from the Welch-processed spectra are plotted in Fig.~\ref{fig:3}(c) against the squared modulus of their corresponding amplitudes. A linear regression analysis of the available data provides an estimate of $\beta_{\text{NL}}$.

\begin{figure}[]
\centering
\includegraphics[width=3in]{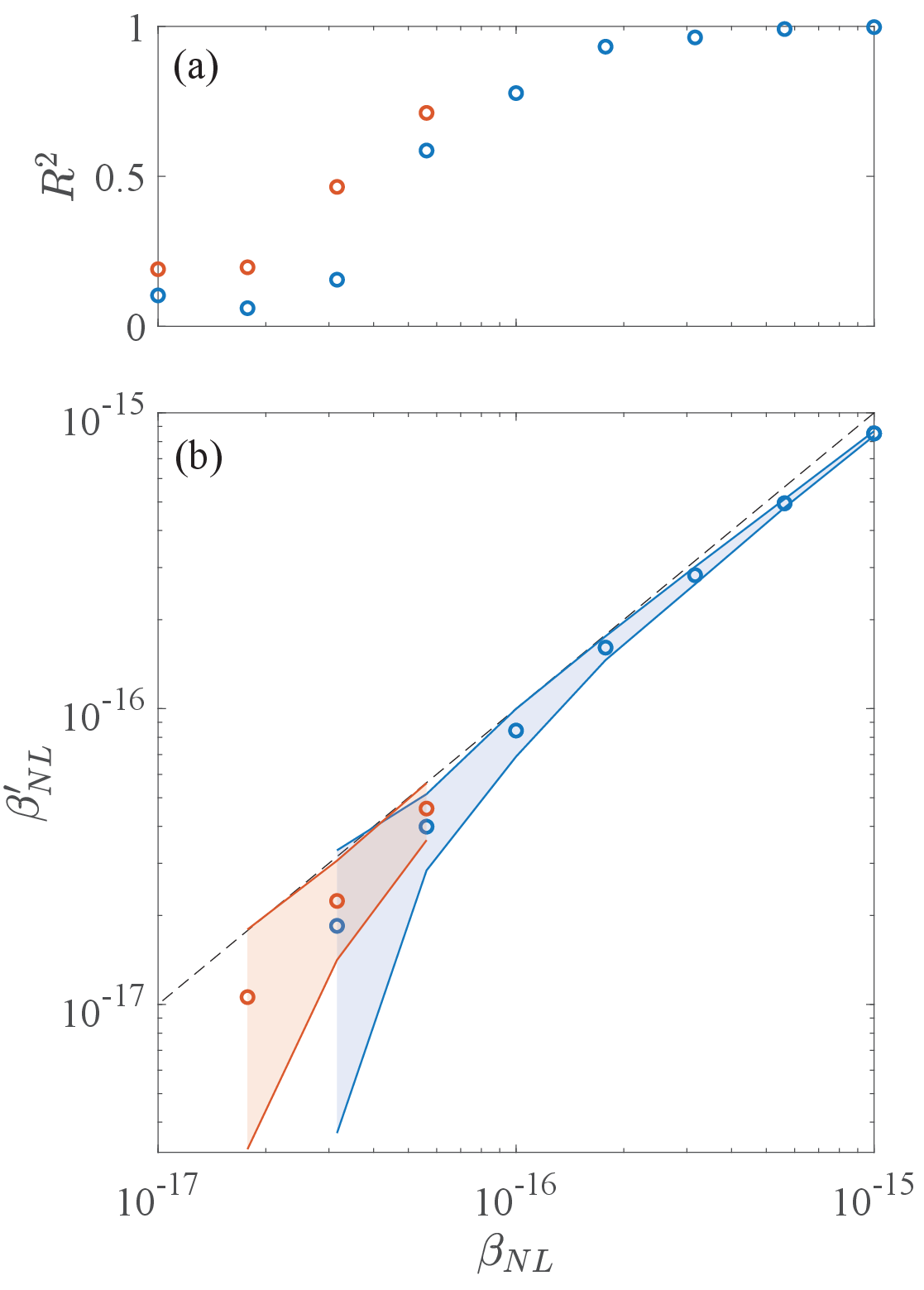}
\caption{(a): The linear regression coefficient $R^2$ of the fitting process for each value of $\beta_{\text{NL}}$. (b): Plot of the estimated $\beta'_{\text{NL}}$ versus $\beta_{\text{NL}}$. Here the blue and red points correspond to the results for $\gamma t=20$, and $\gamma t=60$, respectively. The light areas are the corresponding $95\%$ confidence intervals. Here the parameters are the same as those used in Fig.~\ref{fig:2}
\label{fig:4}}
\end{figure}
The measurement performance of the scheme is characterized in Fig.~\ref{fig:4}. The value of $\beta_{\text{NL}}$ is given as a pre-set parameter ranging from $\beta_{\text{NL}}=10^{-17}$ to $10^{-15}$, and it is subsequently estimated from the noise spectrum to evaluate the resolution and precision of the proposed method. Figure~\ref{fig:4}(a) illustrates that the relationship between linear regression coefficient, denoted by
\begin{equation}
\begin{split}
R^2=1-\frac{\sum_{i=1}^n (\omega'_{b,i} - {\omega''_{b_,i}})^2}{\sum_{i=1}^n (\omega'_{b,i} - \bar{\omega'_{b,i}})^2},
\label{eq:R2}
\end{split}
\end{equation}
versus $\beta_\text{NL}$. For $\beta_{\text{NL}}\geq 10^{-16}$, we have $R^2\simeq 1$, indicating a nearly perfect linear relationship between the two variables, which corresponds to high measurement precision. Near $\beta_{\text{NL}} = 10^{-16.5}$, however, the linear regression coefficient falls below the threshold of $R^2=0.1$. At this critical value, the tiny $\beta_{\text{NL}}$ renders the linear relationship unobservable within the limited frequency resolution, thereby compromising the scheme's practical utility. Consequently, the overall resolution limit is established at a value of $\beta_{\text{NL,lim}}=10^{-16.5}$. Figure~\ref{fig:4}(b) shows the estimated $\beta'_{\text{NL}}$ with varied pre-set $\beta_{\text{NL}}$, with a range from the limit resolution $10^{-16.5}$ to $10^{-15}$, in conjunction with the corresponding $95\%$ confidence intervals.

It is worth noting that the preceding analysis utilizes a time-domain interval of $\gamma t = 20$. The performance of the measuring device can be enhanced through the extension of this interval. We substitute $\beta_{\text{NL}}$, which is located in proximity to the resolution, into Eq.~\eqref{eq:CLE} and subsequently re-simulate the data to $\gamma t=60$.  The generated data is subsequently reprocessed with the stipulated procedure. The corresponding results are then plotted as red points in Fig.~\ref{fig:4}, which shows enhanced regression coefficients $R^2$ at constant $\beta_{\text{NL}}$ values. Consequently, the resolution limit is extended to $\beta_{\text{NL,lim}}=10^{-16.75}$. Given that the scheme is predicated on the stationary dynamics instead of the non-stationary,  it is always feasible to further minimize the measurement limit through further temporal extension.

\subsection{Actual situation: The disturbance caused by parameter difference}
\label{Actual situation}
\begin{figure}[]
\centering
\includegraphics[width=3in]{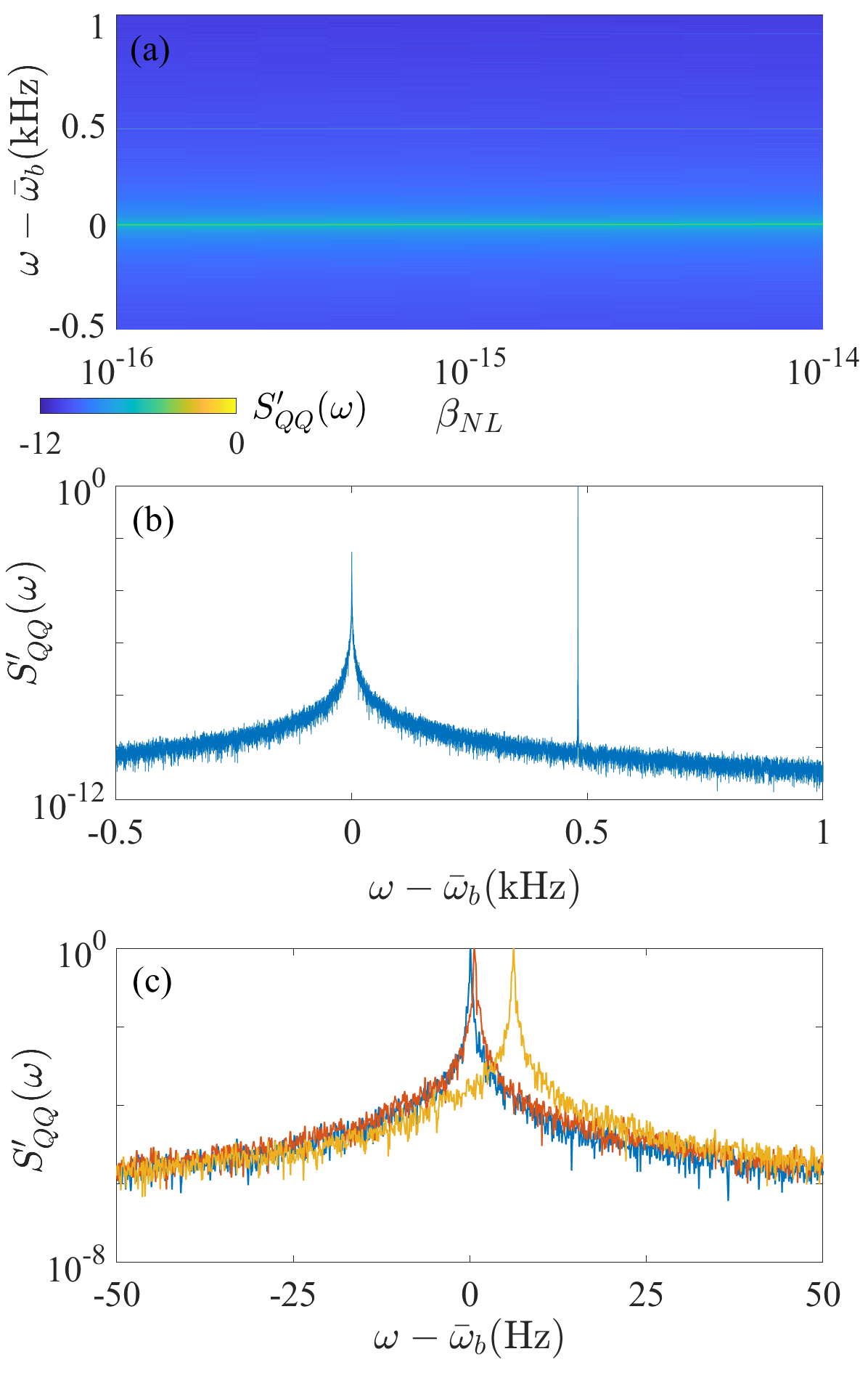}
\caption{(a): The normalized noise spectrum obtained by the Welch's method of the dark mode in the frequency domain with varied per-set GUP parameter $\beta_{\text{NL}}$. (b): The normalized noise spectrum corresponding to $\beta_{\text{NL}}=10^{-16}$. (c): Local amplification near the noise spectrum corresponding to  $\beta_{\text{NL}}=10^{-16}$ (blue),  $\beta_{\text{NL}}=10^{-15}$ (red) and  $\beta_{\text{NL}}=10^{-14}$ (yellow). Here we set $\delta=$\SI{1}{Hz}, $\Delta_2=$\SI{525.48}{kHz} and $P_h=$\SI{100}{\micro\watt}. The other parameters are the same as those used in Fig.~\ref{fig:2}. 
\label{fig:5}}
\end{figure}
As it was mentioned in the associated discussions corresponding to Eqs.~\eqref{eq:bright mode} and~\eqref{eq:dark mode}, in the case that the intrinsic properties of the two oscillators, including their eigenfrequency and Q factor, are not identical, the dark mode remains decoupled from the optical mode but interacts with the bright mode ($\mu\neq 0$). In the event that the dynamics of the bright mode are characterised by a single-mode coherent oscillation, it can be posited that, for the dark mode, the bright mode can be considered as equivalent to a coherent drive, thereby inducing oscillation in the dark mode. The corresponding peak will overwhelm the noise spectrum, which means that a detuning amount must be made between the oscillation frequency of the bright mode and the eigenfrequency of the dark mode,  expressed as $\text{Im}(\Gamma_b)=(\Delta\omega_{m1}+\Delta\omega_{m2})/2=\Delta_p$, so that their peaks are staggered in the frequency domain. The value of $\Delta_p$ depends on the excitation associated with the dark mode. Excessive excitation can result in a scenario where, under specific conditions of broadening, the component of the frequency corresponding to the dispersion to the noise spectrum can exceed the capacity of the noise spectrum. Consequently, in such cases, the detuning amount must increase further, and vice versa. It is noteworthy that, given the identical eigenfrequency of bright mode and dark mode, the detuning between the drive and the dark mode is equivalent to the detuning between the drive and the bright mode. It can thus be concluded that the amplitude of the bright mode will undergo the following transition:
\begin{equation}
\begin{split}
\vert A_b\vert\simeq\dfrac{\sqrt{2}g\mathcal{F}_2}{\gamma}\rightarrow \vert A_b\vert\simeq\dfrac{\sqrt{2}g\mathcal{F}_2}{\Delta_p}.
\label{eq:amplitude de}
\end{split}
\end{equation}
Consequently, when taking the inter-oscillator differences into consideration, it becomes apparent that a greater laser power is required in order to maintain a substantial amplitude of the bright mode and thereby amplify the effect of GUP.

The measurement process is then re-simulated with a frequency difference between the oscillators. Here we still assume that both oscillators have the same dissipation rate $\gamma$ to facilitate the discussion of the acquisition time. This assumption maintains generality, since both frequency differences and dissipation differences can be attributed to the same physical mechanism, that is, induce coupling between bright and dark modes. Specifically, we select $\delta=$\SI{1}{Hz} and $\Delta_2=$\SI{525.48}{kHz}, which results in a \SI{480}{Hz} offset between the frequency of the forced oscillation frequency and the eigenfrequency. The pump intensity is increased to $P_h=$\SI{100}{\micro\watt}, aligning with the previously defined maximum laser power. In Fig.~\ref{fig:5}(a), we show the spectrum of the dark mode as a function of the preset GUP parameter $\beta_{\text{NL}}$. In this instance,  the spectrum exhibits a bimodal structure. In addition to the Lorentz-type noise spectrum, the presence of a delta-function-like spectrum is observed, indicating that the dark mode has been excited to oscillate. In Fig.~\ref{fig:5}(b), we plot the spectrum of the dark mode for $\beta_{\text{NL}}=10^{-16}$. Fig.~\ref{fig:5}(c) zooms in on the noise peak from Fig.~\ref{fig:5}(b), with comparisons to the cases $\beta_{\text{NL}}=10^{-15}$ and $\beta_{\text{NL}}=10^{-14}$, demonstrating that the GUP effect still altering the noise-spectrum frequency, even under mismatch conditions.
\begin{figure}[]
\centering
\includegraphics[width=3in]{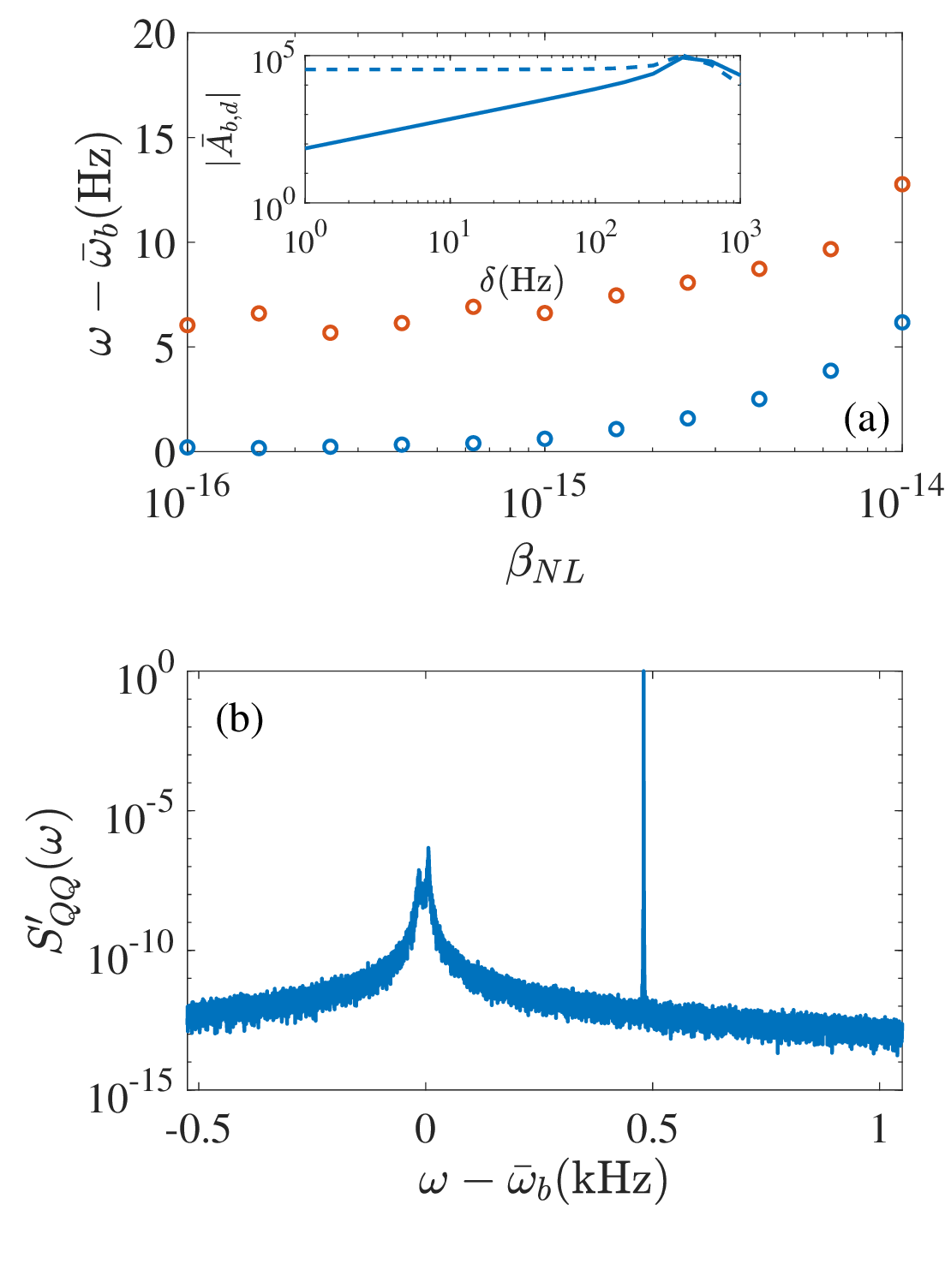}
\caption{(a): The position of the noise spectrum peak obtained by the Welch's method as a function of varied pre-set GUP parameter $\beta_{\text{NL}}$. Blue (red) points correspond to $\delta=$\SI{1}{Hz} (\SI{10}{Hz}). The inset illustrates the average amplitude of the bright and dark mode with varied frequency difference between oscillators $\delta$. (b): The normalized noise spectrum corresponding to $\delta=$\SI{10}{Hz}.  Here the other parameters are the same as those used in Fig.~\ref{fig:5}.
\label{fig:6}}
\end{figure}

Figure~\ref{fig:6} illustrates the noise-peak position as a function of the pre-set $\beta_{\text{NL}}$ in the case that $\delta=$\SI{1}{Hz} and $\delta=$\SI{10}{Hz}, respectively. Owing to the bright-mode amplitude $\vert \bar{A}_b\vert\simeq 3\times 10^4$ being $50\%$ lower than the case that depicted in Fig.~\ref{fig:4}, the limiting resolution degrades to approximately $\beta_{\text{NL,lim}}=10^{-15}$ with the  data acquisition time being sustained at $\gamma t = 20$. This constraint can be alleviated by extending the data acquisition time.  As $\delta$ increases, the dark mode becomes more strongly excited, as illustrated in the inset of Fig.~\ref{fig:6}. For $\delta>$\SI{10}{Hz}, the dark-mode amplitude rises sharply, approaching levels comparable to that of the bright mode. Fig.~\ref{fig:6}(b) illustrates the dark mode spectrum under a larger oscillator frequency difference $\delta$=\SI{10}{Hz}. The noise peak remains unswamped owing to the sufficiently large detuning $\Delta_p$. From Eq.~\eqref{eq:dark mode}, we can obtain
\begin{equation}
\begin{split}
\vert \langle \bar{A}_d\rangle\vert\simeq\left\vert\dfrac{\mu}{\Gamma_d} \langle \bar{A}_b\rangle\right\vert\simeq \dfrac{\delta}{\Delta_p}\vert \langle \bar{A}_b\rangle\vert.
\label{eq:dark mode Amp}
\end{split}
\end{equation}
Given $\delta\ll \Delta_p$, the dark mode amplitude remains substantially smaller than that of the bright mode, as illustrated in the left portion of the inset. Assuming a critical value, $\vert A\vert_{\text{lim}}$, that avoids submerging the noise peak, the GUP induced frequency shift on a single oscillator amplifies to $\vert A\vert^2_{\text{lim}}\beta_{\text{NL}}$. In the dark mode detector, the frequency shift is also amplified by the bright mode resulting in magnification reaching $({\Delta_p}/{\delta})^2\vert A\vert^2_{\text{lim}}\beta_{\text{NL}}$. Comparing the two reveals that the resolution is enhanced by a factor $({\Delta_p}/{\delta})^2$. This is another advantage of dark mode detectors.

\section{Discussion and conclusion}
\label{Discussion and conclusion}
In summary, we propose a scheme for probing GUP at low energy scales by using a full quantum OMS detector composed of two oscillators coupled to a common cavity. The interference between these oscillators generates bright and dark modes, with the dark mode decoupled from the optical field. We demonstrate that the GUP effect induces a frequency shift in the noise spectrum of the dark mode, with the magnitude of the shift depending on the amplitude of the bright mode. Consequently, the GUP parameters can be quantitatively determined by measuring the dark mode's noise spectrum when the entire system is driven by a modulated laser and reaches a stationary oscillation. Relative to approaches relying on nonstationary dynamics of a single oscillator, our scheme offers key advantages: (1).~Replacing transient dissipation with stationary oscillation dynamics removes time-domain limitations imposed by the quality factor, enabling theoretically infinite frequency resolution; (2).~The optical manipulation and excitation are confined to the bright mode, ensuring the dark mode remains unperturbed by radiation pressure, without requiring precise modulation to mitigate cavity-induced frequency corrections. Using parameters akin to those in prior experiments, simulations indicate that the measurement resolution of our scheme reaches $\beta_{\text{NL,lim}}=10^{-16.5}$~($10^{-16.75}$),  by prolonging the data acquisition time to $\gamma t = 20$~($60$), and the corresponding  $\beta_{\text{0,lim}}< 10^{24}$ is $10$ orders of magnitude lower than the electroweak scale and $4$ orders of magnitude lower than the limit reported in Ref.~\cite{Li2025}. In cases of oscillator mismatch, the dark mode is found to be excited, necessitating a detuning adjustment between the oscillation frequency of the bright mode and the eigenfrequency of the dark mode. This trade-off reduces the achievable bright mode amplitude at fixed driving power, resulting in a certain degree of resolution degradation. Nevertheless, it does not fundamentally alter the advantages inherent in our scheme.

The proposed scheme obviates the need for stringent control or precise measurement of parameters, such as inter-oscillator differences, driving strength, and frequency. It can be readily implemented on previously reported experimental platforms~\cite{Sheng2020,Piergentili2021,Piotrowski2023,Cao2025,Piergentili2018,Wu2023,Lake2020}. The approach extends beyond optomechanics, applying to a broad range of systems supporting dark modes. For example, Nanomechanical resonators~\cite{Bagheri2009} and cavity magnomechanics~\cite{Li2018} can also serve as detectors. Moreover, dark-mode detection holds potential for sensing other subtle effects, such as tiny spacetime curvature~\cite{Yang2013} and quantum collapse noise~\cite{Li2016}. Recently, the quality factor of the oscillator has been optimized to $\text{Q}\sim 10^8$~\cite{Marzioni2025}, suggesting that the accuracy of our scheme can be further improved.
 
\begin{acknowledgements}
W.~L. is supported by the National Natural Science Foundation of China (Grants No.~12304389), the Scientific Research Foundation of NEU (Grant No. 01270021920501*115). N.~E.~S. acknowledges financial support from NQSTI within PNRR MUR Project PE0000023-NQSTI. W.~Z. is supported by the National Natural Science Foundation of China (Grants No.~12074206 and 22103043). J.~C. is supported in part by the National Natural Science Foundation of China (Grants No.~11704205). 
\end{acknowledgements}

\end{document}